\documentclass[%
 aip,
 apl,
 amsmath,amssymb,
 reprint,%
]{revtex4-1}

\usepackage{graphicx}
\usepackage{dcolumn}
\usepackage{bm}

\usepackage[utf8]{inputenc}
\usepackage[T1]{fontenc}
\usepackage{accents}
\usepackage{mathptmx}
\usepackage{etoolbox}
 
\usepackage[colorlinks=true, allcolors=blue]{hyperref}
\usepackage[dvipsnames]{xcolor}

\definecolor{linkcolor}{HTML}{0176ba}
\definecolor{urlcolor}{HTML}{0176ba} 
\definecolor{citecolor}{HTML}{900020}
\hypersetup{pdfstartview=FitH,  linkcolor=linkcolor,urlcolor=urlcolor, citecolor=citecolor, colorlinks=true}
\usepackage{comment}
\makeatletter
\def\@email#1#2{%
 \endgroup
 \patchcmd{\titleblock@produce}
  {\frontmatter@RRAPformat}
  {\frontmatter@RRAPformat{\produce@RRAP{*#1\href{mailto:#2}{#2}}}\frontmatter@RRAPformat}
  {}{}
}%
\makeatother

\begin{document}


\title{Optimal multipole center for subwavelength acoustic scatterers}

\author{N. Ustimenko}
\affiliation{Institute of Theoretical Solid State Physics, Karlsruhe Institute of Technology, Kaiserstrasse 12, Karlsruhe, D-76131, Germany}%

\author{C. Rockstuhl}
\affiliation{Institute of Theoretical Solid State Physics, Karlsruhe Institute of Technology, Kaiserstrasse 12, Karlsruhe, D-76131, Germany}%
\affiliation{Institute of Nanotechnology, Karlsruhe Institute of Technology, Kaiserstrasse 12, Karlsruhe, D-76131, Germany}

\author{A. V. Kildishev}
\email[Authors to whom correspondence should be addressed: ]{kildishev@purdue.edu}
\affiliation{Elmore Family School of Electrical and Computer Engineering and Birck Nanotechnology Center, Purdue University, 1205 W State St, West Lafayette, IN 47907, USA}%


\begin{abstract}
The multipole expansion is a powerful framework for analyzing how subwavelength-size objects scatter waves in optics or acoustics. The calculation of multipole moments traditionally uses the scatterer's center of mass as the reference point. The theoretical foundation of this heuristic convention remains an open question. Here, we challenge this convention by demonstrating that a different, optimal multipole center can yield superior results. The optimal center is crucial -- it allows us to accurately express the scattering response while retaining a minimum number of multipole moments. Our analytical technique for finding the optimal multipole centers of individual scatterers, both in isolation and within finite arrays, is validated through numerical simulations. Our findings reveal that such an optimized positioning significantly reduces quadrupole contributions, enabling more accurate monopole-dipole approximations in acoustic calculations. Our approach also improves the computational efficiency of the T-matrix method, offering practical benefits for metamaterial design and analysis.
\end{abstract}

\maketitle


The efficient modeling of sound scattering by finite heterogeneous arrays of compact objects (scatterers) varying in shape and material properties represents a fundamental challenge in acoustics. 
This problem has gained renewed attention with the advent of modern high-performance computing, which has enabled the practical implementation of transition matrix (T-matrix) schemes.~\cite{Kinsler1999Dec, Sainidou2005Mar, Gong2017Jan, Gong2019Jun, Ganesh2022Mar,Yang2023Jun,Li2024Aug} 
Contemporary T-matrix frameworks can accurately capture complex interactions among arrayed scatterers, even when the individual scatterers exhibit structural and material diversity.
This capability becomes particularly critical in the emerging field of inverse-designed metamaterials~\cite{Haberman2016Jun,Wang2016Nov,Bertoldi2017Oct,Sieck2017Sep,Colton2019,Ronellenfitsch2019Sep,Krishna2022Aug,Long2022Oct,Li2023Mar,Fang2024Jun} and their physics-based optimization algorithms using T-matrix solvers.~\cite{Zhan2018Feb,Zhelyeznyakov2020Jan,Ustimenko2021Oct,Gladyshev2023Sep,Garg2024Sep} The classical~\cite{ Clebsch1863Jan,Mie1908Jan,Lorenz2019Aug}\textsuperscript{,}\footnote{Ref.~[\onlinecite{Lorenz2019Aug}] above has been translated from the original Danish text Ref.~[\onlinecite{Lorenz1890Dan}] by Jeppe Revall Frisvad, DTU, and Helge Kragh, U. of Copenhagen, Denmark}\nocite{Lorenz1890Dan} and more recent studies~\cite{Muhlig2011Jun,Fernandez-Corbaton2015Dec,Evlyukhin2016Nov,Smirnova2016Nov,Zhu2016May,Alaee2018Jan,Alaee2019Jan,Kovacevich2021Oct,Riccardi2022Sep} on multipole expansion, a versatile method for modeling scattering by subwavelength scatterers in optics and acoustics, form the basis of these approaches.  

In the T-matrix method, the incident and scattered fields for any compact scatterer are expressed as a truncated multipole expansion. The selection of the spatial coordinate relative to which these fields are expanded is critical to the accuracy of the method.
Frequently, the multipoles are positioned at the scatterer's center of mass.~\cite{evlyukhin2011multipole} 
However, the multipole moments induced in a scatterer by incident fields depend on the chosen expansion center, and generally, no unique choice exists.~\cite{evlyukhin2011multipole,Evlyukhin2013Oct} An arbitrary coordinate center can be selected as the expansion center, but it requires retaining an increasing number of multipoles in the expansion to express the scattering response accurately. Computationally, this quickly becomes extremely demanding, if not impossible. Therefore, identifying an optimal center is critical to minimize the number of multipole terms retained in the expansion.  

\begin{figure}
    \centering
    \includegraphics[scale=0.33]{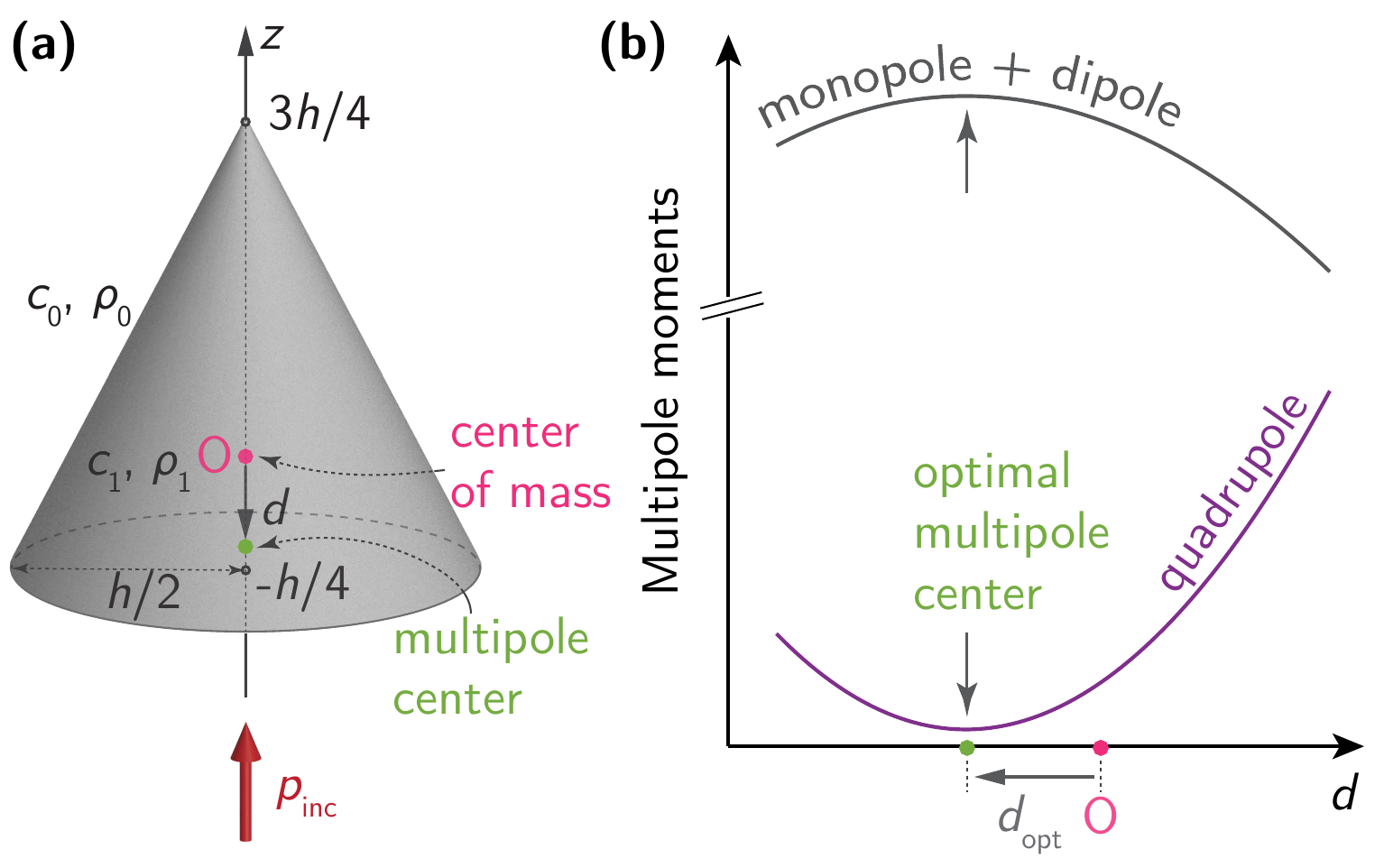}
    \caption{Sketch of the considered problem. (a) A cone of height $h$ under insonification of an acoustic plane wave propagating along the $z$-axis. The center of the multipole expansion (green point) is offset from the center of mass (pink point $O$, $z = 0$) by a distance $d$. (b) Squared amplitudes of first three multipole moments of the cone vs. $d$ at a fixed frequency. The quadrupole moment exhibits a minimum, whereas the sum of the monopole and dipole moments has a maximum, manifesting the optimal multipole center's position $z = d_{\rm opt}$.}
    \label{fig:sketch}
\end{figure}

Our study addresses this challenge in acoustics by seeking to minimize the number of multipoles retained in the expansion when describing the acoustic scattering response from either scatterers or finite arrays thereof. Through both analytical and numerical approaches, we examine the scattering of acoustic pressure waves by subwavelength scatterers to determine the location of the optimal multipole center (OMC). We demonstrate that selecting this optimal center minimizes the quadrupole contribution, creating ideal conditions for applying the monopole-dipole approximation (MDA) in acoustics widely used in the literature.~\cite{Silva2014Nov,Toftul2019Oct,Wei2020Aug,Long2020Jun,Smagin2023Oct} This approach can be applied to both isolated scatterers and compact arrays where simple multipole sources interact at short distances. 

We begin with a mathematical formulation of our problem. The scattered pressure $p_{\rm sca}(k; \mathbf{r})$ of an acoustic scatterer can be expressed as a truncated expansion in scalar spherical waves (multipoles) weighted by multipole moments~\cite{Blackstock2000Apr}
\begin{align}
\label{eq_psca}
    p_{\rm sca}(k; \mathbf{r}) = \sum_{\ell = 0}^{\ell_{\rm max}} \sum_{m = -\ell}^{\ell} p_{\ell, m}(k; \mathbf{d}) \psi^{(3)}_{\ell, m}(k; \mathbf{r} - \mathbf{d})\,.
\end{align}
$\ell$ and $m$ is the degree and order of the singular multipole $\psi^{(3)}_{\ell, m}(k; \mathbf{r} - \mathbf{d})$ [see Sec.~S1A in Supplementary material (SM) for its definition]. $\mathbf{r} \neq \mathbf{d}$, and $k = \omega/c_0$ is the wavenumber in the surrounding medium. $\mathbf{d}$ is an offset of the multipole center relative to the scatterer's center of mass defined as $\mathbf{r}_{\rm CM} = \bm{0}$ [point $O$ in Fig.~\ref{fig:sketch}(a)]. For Eq.~\eqref{eq_psca} to be valid, the coordinate $\mathbf{r}$ must belong to the space region outside a sphere circumscribing the scatterer.~\cite{Auguie2016May} We aim to choose the optimal $\mathbf{d}_{\rm opt}$ that simultaneously provides the best precision in calculating physical quantities and the lowest maximal multipole degree $\ell_{\rm max}$ in the expansion. In this work, we assume that the offset $\mathbf{d}$ is subwavelength, that is, $k|\mathbf{d}|\ll1$. 

To determine $\mathbf{d}_{\rm opt}$, we need to know the dependence of the multipole moments on $\mathbf{d}$. Using the translation coefficients of spherical waves $ \alpha^{(1)}_{\ell m \ell' m'}(k; \mathbf{d})$ [Sec.~S1C in~SM],~\cite{Martin2006Aug} we link the multipole moments obtained for an expansion center shifted to $\mathbf{d}$ to the multipole moments obtained at the center of mass, {\it i.e.}, at $\mathbf{d} = \bm{0}$ [see Sec.~S1D in SM]
\begin{align}
\label{eq_coeffs_shifted}
    p_{\ell, m}(k; \mathbf{d}) = \sum_{\ell' = 0}^{\ell_{\rm max}} \sum_{m' = -\ell'}^{\ell'} \alpha^{(1)}_{\ell m \ell' m'}(k; \mathbf{d}) p_{\ell', m'}(k; \bm{0})\,.
\end{align}
$p_{\ell', m'}(k; \bm{0})$ can be computed using, {\it e.g.}, the acoustic T-matrix of the scatterer~\cite{Waterman1969Jun,Waterman2009Jan} [see Sec.~(S6) in SM], or upon post-processing simulations from a full-wave solver (see, {\it e.g.}, Appendix B in Ref.~[\onlinecite{Tsimokha2022Apr}], and Ref.~[\onlinecite{Deriy2022Feb}]). Equation~\eqref{eq_coeffs_shifted} is of significant practical importance -- it shows that to obtain the entire dependence of the multipole moments on $\mathbf{d}$, we \textit{can calculate the multipole moments only once}, if the chosen $\ell_{\rm max}$ is sufficient.

Further, we verify our semi-analytical approach in two examples. As a first example, we find the OMC for a cone-shaped scatterer with cylindrical symmetry along the $z$-axis [see Fig.~\ref{fig:sketch}(a)]. The symmetry is conserved for a plane wave excitation along the $z$-axis, and only the zonal multipoles (with $m = 0$) are excited. Moreover, the offset along the $z$-axis $\mathbf{d} = d \hat{\mathbf{z}}$ affects only the zonal multipole content of the scatterer because the shifted system remains axisymmetric. Furthermore, we consider a spectral range where the first three multipoles sufficiently approximate the scattering for all $d$ we examine, {\it i.e.}, $\ell_{\rm max} = 2$ can be held. In this case, the scattering cross-section of the cone is $\sigma_{\rm tot}(k) = k^{-2} \sum_{\ell=0}^{\ell_{\rm max}}|p_{\ell, 0}(k; \mathbf{d})|^2$. Here, we consider the offsets $d$ that are sufficiently small to make $\sigma_{\rm tot}(k)$ almost independent on $\mathbf{d}$, while its MDA $\sigma_{\rm md}(k;\mathbf{d}) = k^{-2} \sum_{\ell=0}^{1}|p_{\ell, 0}(k;\mathbf{d})|^2$ still depends on $\mathbf{d}$. Therefore, we define $\mathbf{d}_{\rm opt}$ as $\min_{\mathbf{d}}|p_{2, 0}(k; \mathbf{d})|^2$ for a given $k$ to reduce error $|\sigma_{\rm md}(k;\mathbf{d}) - \sigma_{\rm tot}(k)|$. We then compute $\mathbf{d}_{\rm opt}$ from equation $\bm{\nabla}_{\mathbf{d}} |p_{2, 0}(k; \mathbf{d})|^2 = 0$.

Although standard numerical methods~\cite{Chong2013} can be used to find $\mathbf{d}_{\rm opt}$, we solve this problem analytically. We start by adapting~\eqref{eq_coeffs_shifted} to the considered case and write the off-origin quadrupole moment as a function of a normalized offset $(kd)$. For all zonal multipole coefficients, we make substitutions $p_{2}(kd) \equiv p_{2, 0}(k; d\hat{\mathbf{z}})$, and $p_{\ell} \equiv p_{\ell, 0}(k; \bm{0})$ for brevity, and then arrive at the following compact equation
\begin{align}
\label{eq_quadrupole_transl}
\begin{aligned}
    p_{2}(kd) = p_2 + \mathbf{A}_2(kd)\cdot \mathbf{p}\,,
\end{aligned}
\end{align}
with $\mathbf{p}=\begin{bmatrix}
p_0, 
p_1, 
p_2
\end{bmatrix}^{\rm T}
$, $\mathbf{A}_2(kd) =
\begin{bmatrix}
\alpha_0(kd),
\alpha_1(kd),
\alpha_2(kd)
\end{bmatrix}^{\rm T}
$, and $\alpha_{\ell'}(kd)\equiv\alpha^{(1)}_{2 0 \ell' 0}(k; d\hat{\mathbf{z}})-\delta_{2,\ell'}$. The long-wavelength approximation (LWA) $kd \ll 1$ of the spherical Bessel functions [see Sec.~S1B in SM] gives an approximate separation vector $\widetilde{\mathbf{A}}_2(kd) =
\begin{bmatrix}
\widetilde{\alpha}_0(kd),
\widetilde{\alpha}_1(kd),
\widetilde{\alpha}_2(kd)
\end{bmatrix}^{\rm T}
$
\begin{align}
\label{eq_transl_coeffs_expansion}
\begin{aligned}
\widetilde{\mathbf{A}}_2(kd) 
=
\begin{bmatrix}
\frac{1}{3\sqrt{5}} (kd)^2, 
-\frac{2}{\sqrt{15}}(kd), 
-\frac{11}{42}(kd)^2
\end{bmatrix}^{\rm T}\,,
\end{aligned}
\end{align}
where the tilde is used to distinguish the accurate values of translation coefficients $\alpha_{\ell'}$ from their LWA $\widetilde{\alpha}_{\ell'}$. 
After substituting~\eqref{eq_transl_coeffs_expansion} into~\eqref{eq_quadrupole_transl}, we arrive at a quadratic approximation of $|p_{2}(kd)|^2$
\begin{align}
\label{eq_quadrupole_interpol}
    |\tilde{p}_{2}(kd)|^2 = f_{0,k} + f_{1,k} (kd) + f_{2,k} (kd)^2\,,
\end{align}
where $f_{0,k} = |p_{2}|^2$, $f_{1,k} = -\frac{4}{\sqrt{15}}\Re\left[ p_{1}^* p_{2}\right]$, and $f_{2,k} = \frac{4}{15}|p_{1}|^2 - \frac{11}{21}|p_{2}|^2 + \frac{2}{3\sqrt{5}}\Re\left[ p_{0}^* p_{2}\right]$ with a real $k$ [for a complex $k$, we refer to Sec.~S2C in SM] and $p_{\ell} = p_{\ell}(k)$, calculated for $d = 0$. The accuracy of Eqs.~\eqref{eq_quadrupole_transl} and~\eqref{eq_quadrupole_interpol} in the considered spectral range is confirmed in Sec.~S2A in SM. Similar interpolations can be derived for the coefficients $p_{0}(kd)$ and $p_{1}(kd)$ [see Sec.~S2B in SM] as well as for high-degree terms. The main outcome of~\eqref{eq_quadrupole_interpol} is an analytical formula for the optimal offset being a solution to the equation $\partial_{(kd)} |\tilde{p}_{2}(kd)|^2 = 0$ ($k$ is fixed) that provides $d_{\rm opt} = -f_{1,k}/\left(2 k f_{2,k}\right)$, or
\begin{align}\label{eq_dopt_formula}
    d_{\rm opt}= \frac{\sqrt{3}}{k} \frac{\Re\left[ p_{1}^* p_{2}\right]}{\frac{2}{\sqrt{5}}|p_{1}|^2 - \frac{11 \sqrt{5}}{14}|p_{2}|^2 + \Re\left[ p_{0}^* p_{2}\right]}\,,
\end{align}
where the sign of second derivative $f_{2,k} > 0$ and $k \neq 0$ have been implied.
Thus, once the multipole moments at $d = 0$ are known, the OMC's position is directly given by~\eqref{eq_dopt_formula}. Moreover, ~\eqref{eq_dopt_formula} can be used even if the initial multipole center is at an arbitrary point $d = d_0$. Then, a translation $\left(d_{\rm opt} + d_0\right)$ gives the OMC.

To increase the accuracy, one can extend expansion~\eqref{eq_quadrupole_interpol} up to the next even power of $(kd)$,
\begin{align}
\label{eq_quadrupole_interpol_quartic}
    |\breve{p}_{2}(kd)|^2 =\sum_{n=0}^{4} f_{n,k}(kd)^n,
\end{align}
where $f_{3,k}$ and $f_{4,k}$ are given in Sec.~S2C in SM. In this case, an optimal offset is calculated as a pure real solution to the cubic equation $\partial_{(kd)} |\breve{p}_{2}(kd)|^2 = 0$ being
\begin{align}
\label{eq_quadrupole_interpol_quartic_grad}
f_{1,k} + 2 f_{2,k} (kd) + 3 f_{3,k} (kd)^2 + 4 f_{4,k} (kd)^3 = 0.
\end{align}

\begin{figure}[h!]
    \centering \includegraphics[scale=0.75]{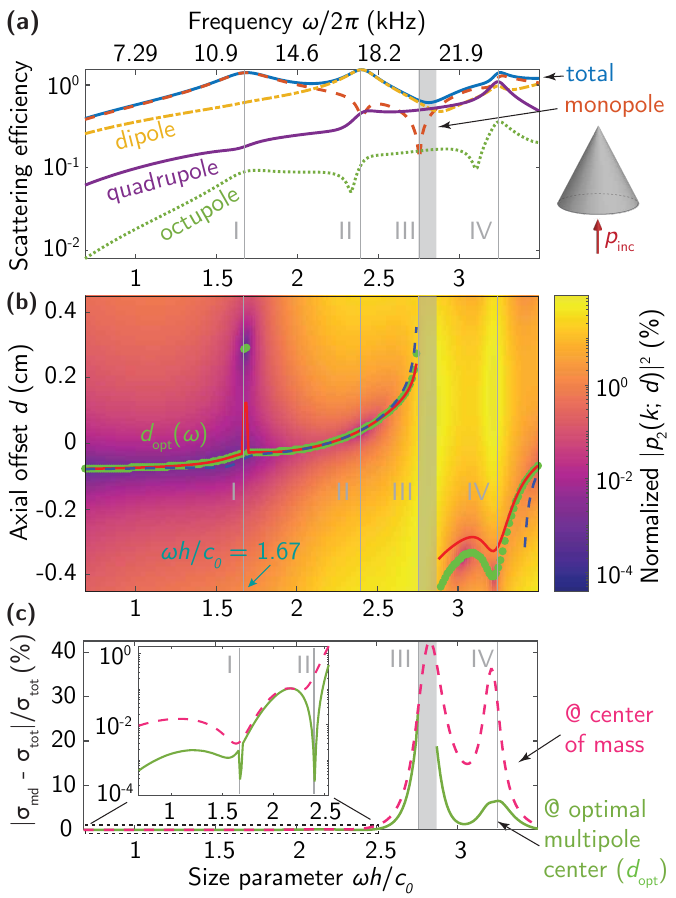}
    \caption{(a) Scattering efficiency $\sigma_{\rm tot}(k)/(0.25 \pi h^2)$ of the cone, shown in the inset, with its multipole decomposition for the multipole center at $d = 0$. (b) Normalized zonal quadrupole moment of the cone $|p_{2}(k;d)|^2 / \sum_{\ell \neq 2} |p_{\ell}(k;d)|^2$ for $\ell_{\rm max} = 5$. The green dots show the values of $d_{\rm opt} = \min_{d}|p_{2}(k;d)|^2$, obtained directly from the full-wave numerical simulations. The dashed blue and solid red lines show $d_{\rm opt}$ calculated with the quadratic [Eq.~\eqref{eq_dopt_formula}] and quartic [Eq. ~\eqref{eq_quadrupole_interpol_quartic_grad}] approximations of $|p_2(k;d)|^2$, respectively. (c) Error of the scattering cross-section for $\ell_{\rm max} = 1$ relative to that for $\ell_{\rm max} = 5$ at $d = 0$ (pink dashed) and $d = d_{\rm opt}$, calculated in the full-wave model (green solid). Inset: the same but on a logarithmic scale for $\omega h / c_0 \leq 2.5$.}
    \label{fig:quadrupole_vs_dcm_vs_freq}
\end{figure}

To validate the analytical results, we perform a numerical simulation of the scattering of a plane wave by a cone of height $h = 1$~cm and radius of $h /2$. The cone is made of a material with the speed of sound $c_1$ and mass density $\rho_1$. It is placed in a medium with $c_0/c_1 = \sqrt{21}$ and $\rho_0/\rho_1 = 1/7$ [see Fig.~\ref{fig:sketch}(a)]. We use our in-house T-matrix-based code \textit{acoustotreams},~\cite{acoustotreams} which is an adaptation of the electromagnetic scattering code \textit{treams} for acoustic scattering.~\cite{Beutel2024Apr} The acoustic T-matrix of the cone was calculated for $d = 0$ in a spectral range between 5-25.5 kHz with a step of 0.1 kHz using the finite-element method (Pressure Acoustics interface in \textsc{COMSOL Multiphysics}, v.~6.2).~\cite{comsol}
The acoustic T-matrix was then exported to \textit{acoustotreams} to calculate the offset multipole moments~\eqref{eq_coeffs_shifted} with $\ell_{\rm max} = 5$ (full-wave model). 

Figure~\ref{fig:quadrupole_vs_dcm_vs_freq} presents the results of the full-wave numerical simulations, which are the cone scattering efficiency in (a), the normalized quadrupole moment map in (b) with numerical values of $d_{\rm opt}$ shown by green markers, and the relative error for the MDA of the scattering cross-section in (c) calculated for the multipole center placed at the center of mass and OMC (numerical values). Figure~\ref{fig:quadrupole_vs_dcm_vs_freq}(b) also compares the numerical values of $d_{\rm opt}$ to the analytical values in the quadratic~\eqref{eq_dopt_formula} and quartic~\eqref{eq_quadrupole_interpol_quartic_grad} approximations, respectively. Multiple observations can be made.

First, the OMC converges to $d_{\rm opt} = -0.075$~cm in a subwavelength regime $\omega h/c_0 \lesssim 1$ [see Fig.~\ref{fig:quadrupole_vs_dcm_vs_freq}(b)]. Thus, the OMC is located below the center of mass, similar to the magnetic multipole center of electromagnetic scatterers.~\cite{Kildishev2023Sep} 
The OMCs predicted by all three models coincide for $\omega h/c_0 < 2.75$ except at the monopole resonance $\omega_{\rm I} h/c_0 = 1.68$. At this frequency, the quartic approximation $|\breve{p}_2|^2$ fails near the actual minimum of $|p_2|^2$ taking nonphysical negative values. Therefore, the sextic approximation should be used. We do not show it here for brevity, but it can be found in Sec.~(S2.C.4) in SM. For optical scatterers, we have already demonstrated that the optimal multipole center differs from the scatterer's center of mass and is, in fact, dispersive~\cite{Kildishev2023Sep}.
In contrast, at dipole resonance $\omega_{\rm II} h/c_0 = 2.4$, all models predict the same OMC with a significant suppression of the error [see the inset in Fig.~\ref{fig:quadrupole_vs_dcm_vs_freq}(c)]. 
In Fig.~\ref{fig:quadrupole_vs_dcm_vs_freq}(b), we also see that the negative OMC turns into a large positive OMC around $\omega_{\rm III} h/c_0 \simeq 2.75$ (the gray region in Fig.~\ref{fig:quadrupole_vs_dcm_vs_freq}), and the second derivatives of~\eqref{eq_quadrupole_interpol} and~\eqref{eq_quadrupole_interpol_quartic} become negative [see Sec.~S3 in SM]. In this case, the monopole contribution to the scattering is suppressed, while the quadrupole becomes substantial. Figure~\ref{fig:quadrupole_vs_dcm_vs_freq}(c) also confirms that neglecting the quadrupole around $\omega \simeq \omega_{\rm III}$ leads to an error larger than $40$\%. Hence, the OMC should be redefined as a minimum of a higher-degree multipole moment. 
Surprisingly, at the quadrupole resonance $\omega_{\rm IV} h/c_0 = 3.24$, the quadrupole moment can still be neglected with an error of $10$\% since the monopole also exhibits a resonant-like response. Thus, a center for the multipole expansion of the fields placed at the optimal point provides the best conditions for the MDA. 
The error in the scattering cross-section is consistently smaller when the OMC is chosen for the expansion compared to when the center of mass is chosen [see Fig.~\ref{fig:quadrupole_vs_dcm_vs_freq}(c)]. At the same time, one can use Eq.~\eqref{eq_dopt_formula} to predict the positions of the OMC for $\omega h / c_0 \leq 2.75$ when $f_{2,k} > 0$. For higher frequencies, Eq.~\eqref{eq_quadrupole_interpol_quartic_grad} or the higher approximations should be employed [see Fig.~\ref{fig:quadrupole_vs_dcm_vs_freq}(b)].

In a further analysis, we demonstrate that the OMC can be found not only for individual scatterers but also for finite arrangements thereof. The framework of the T-matrix method can be efficiently applied to acoustic multiple scattering problems~\cite{Peterson1975Jan, Kafesaki1999Nov, Psarobas2000Jul}, where one needs to distinguish a formulation in a local and global description of the scattered fields of arrangements~\cite{Suryadharma2017Jul}. In the local description (LD), the fields are expanded at the $N$ local positions of all $N$ scatterers, $\mathbf{r}_i$ with $i = 1..N$. The multiple scattering equation is solved to obtain the self-consistent multipole moments that account for the interactions of the scatterers with the incident field and with each other (see, {\it e.g.},~(17) in Ref.~[\onlinecite{Beutel2024Apr}]). In the LD, the multipole moments are calculated when the multipole centers coincide with the centers of mass of scatterers located at $\mathbf{r}_i$. In the global description (GD), the fields are expanded from all particles using a global expansion center at a single point $\mathbf{d}$. In this regard, the global multipole moments depend on $\mathbf{d}$ and the local multipole moments as
\begin{align}
\label{eq_coeffs_shifted_ms}
 p^{\rm (global)}_{\ell, m}(k; \mathbf{d}) = \sum_{\ell' = 0}^{\ell_{\rm max}} \sum_{m' = -\ell'}^{\ell'} \sum_{i = 1}^N \alpha^{(1)}_{\ell m \ell' m'}(k; \mathbf{d} - \mathbf{r}_i) p_{\ell', m'}^{\rm (local)}(k; \mathbf{r}_i)\,.
\end{align} 

Equation~\eqref{eq_coeffs_shifted_ms} is similar to~\eqref{eq_coeffs_shifted} for the single-particle case, but it includes a sum over the scatterers. The other feature is more inconspicuous. The maximal multipole degree in the GD $L_{\rm max}$ can be larger than in the LD $\ell_{\rm max}$. Thus, the transition from the LD to the GD is reasonable when the size of an arrangement T-matrix in the GD, $(L_{\rm max} + 1)^2$, is smaller than the size of an arrangement T-matrix in the LD, $N(\ell_{\rm max} + 1)^2$. 

\begin{figure}[!ht]
    \centering \includegraphics[scale=0.75]{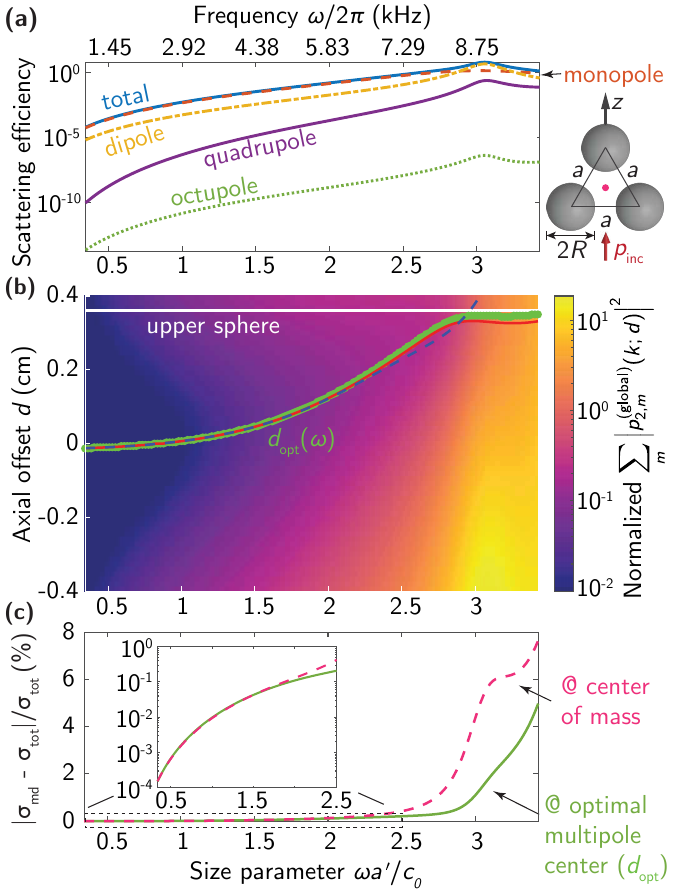}
    \caption{The same as in Fig.~\ref{fig:quadrupole_vs_dcm_vs_freq} but for a trimer of spherical scatterers [see the inset of panel (a)] using the global description~\eqref{eq_coeffs_shifted_ms} of the trimer acoustic response. (a) The scattering efficiency is $\sigma_{\rm tot}(k)/(0.25 \pi [a']^2)$ where $a' = a + 2R$. The pink dot indicates the center of mass. (b) Normalized global quadrupole moment of the trimer $\sum_{m} |p^{\rm (global)}_{2,m}(k;d)|^2 / \sum_{\ell \neq 2,m} |p^{\rm (global)}_{\ell,m}(k;d)|^2$ for $\ell_{\rm max} = 2$ and $L_{\rm max} = 4$. The green dots show the values of $d_{\rm opt} = \min_{d}\sum_m|p^{\rm (global)}_{2,m}(k;d)|^2$, obtained directly from the full-wave numerical simulations. The dashed cyan and solid red lines show $d_{\rm opt}$ calculated with the quadratic [Eq.~\eqref{eq_dopt_formula}] and quartic [Eq.~\eqref{eq_quadrupole_interpol_quartic_grad}] approximations of the zonal term $|p^{\rm (global)}_{2,0}(k;d)|^2$, respectively. (c) Error of the scattering cross-section for $L_{\rm max} = 1$ relative to that for $L_{\rm max} = 4$ at $d = 0$ (pink dashed) and $d = d_{\rm opt}$, calculated in the full-wave model (green solid). Inset: the same but on a logarithmic scale for $\omega a^{\prime} / c_0 \leq 2.5$.}
    \label{fig:trimer_quadrupole_vs_dcm_vs_freq}
\end{figure}

As an example, we consider a trimer of spherical particles shown in Fig.~\ref{fig:trimer_quadrupole_vs_dcm_vs_freq}(a). The spheres of radius $R = 0.5$~cm are placed at the nodes of an equilateral triangle of side $a = 3R$. The material parameters (mass density and speed of sound) of the scatterer material and background are the same as before. Hereafter, we can use the same formalism as before but in a multiple scattering scenario to find the minima of the quadrupole moment in the GD $\sum_{m=-2}^2 |p^{\rm (global)}_{2,m}(k;\mathbf{d})|^2$. An interaction between the spheres can induce the multipoles with $m \neq 0$. However, their contribution is smaller than that of zonal multipoles. Therefore, we can apply Eqs.~\eqref{eq_dopt_formula} and~\eqref{eq_quadrupole_interpol_quartic_grad} to predict $d_{\rm opt}$. We assume the range of frequencies 1--11 kHz where $\ell_{\rm max} = 1$ and $L_{\rm max} = 2$ are sufficient. If we manage to find the OMC where even $L_{\rm max} = 1$ provides good accuracy, then we will decrease the size of the trimer T-matrix from 12 to 4, although one has to track $d_{\rm opt}(\omega)$. 

Figure~\ref{fig:trimer_quadrupole_vs_dcm_vs_freq} presents the same content as Fig.~\ref{fig:quadrupole_vs_dcm_vs_freq} but for the trimer. The main difference lies in the behavior of $d_{\rm opt}$. In the static case, the system is symmetric concerning the rotations by $\pi/3$ and $2\pi/3$ in the $xz$ plane; hence $d_{\rm opt} \to 0$ when $\omega a'/c_0 \to 0$, {\it i.e.}, the OMC coincides with the center of mass of the trimer [see Fig.~\ref{fig:trimer_quadrupole_vs_dcm_vs_freq}(b)]. Here, $\omega a'/c_0$ is an effective trimer size parameter with $a' = a + 2R$. As $\omega a'/c_0$ increases, $d_{\rm opt}$ also increases. From $\omega a'/c_0 = 2.6$, the OMC remains at the same point $d_{\rm opt} \approx 0.35$~cm close to the center of the upper sphere and provides a better accuracy of the MDA [see Fig.~\ref{fig:trimer_quadrupole_vs_dcm_vs_freq}(c)]. Both quadratic and quartic models predict $d_{\rm opt}$ for $\omega a'/c_0 \leq 2.6$ very well, although the quadratic model fails for $\omega a'/c_0 > 2.6$ due to the resonant response of the trimer [see Fig.~\ref{fig:trimer_quadrupole_vs_dcm_vs_freq}(a)]. In Sec.~S4 in SM, we also consider this trimer rotated by $\pi/2$ in the $xz$ plane, for which $\mathbf{d}_{\rm opt}$ is traced as a trajectory, with the initial point at $\mathbf{d}_{\rm opt} = \bm{0}$, as the frequency increases.

\begin{figure}
    \centering \includegraphics[scale=0.63]{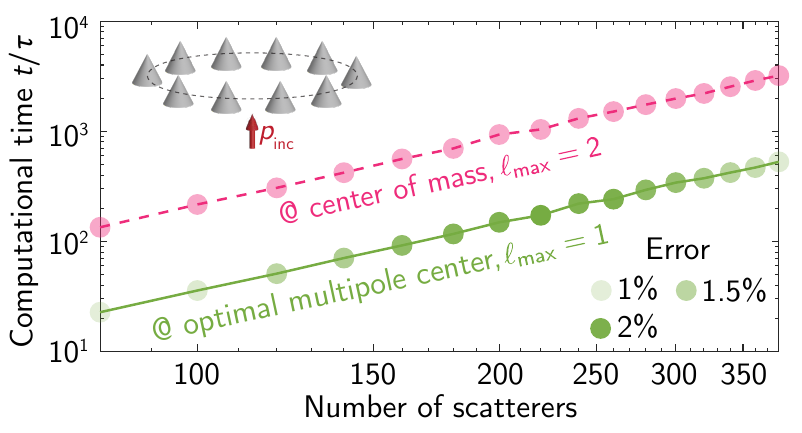}
    \caption{Normalized computational time versus the number of cone-shaped scatterers arranged in a ring depicted in the inset. The expansion center for the single cone is placed at either its center of mass with $\ell_{\rm max} = 2$ (pink dashed) or the OMC with $\ell_{\rm max} = 1$ (green solid). The transparency of the markers indicates the corresponding error of the scattering cross-section.}
    \label{fig:time_vs_N}
\end{figure}
In the end, we demonstrate that the OMC can improve the computational efficiency of the acoustic T-matrix method in multiple scattering problems. For that purpose, we consider a ring of $N$ identical cones (the same as those previously considered) placed at a distance $D = 2.6$~cm between the nearest neighbors. The acoustic T-matrix of the individual cone is calculated at frequency $\omega h/c_0 = 2.5$ for the multipole expansion center placed at $d = 0$ with $\ell_{\rm max} = 2$, and at the OMC with $\ell_{\rm max} = 1$ [see Fig.~\ref{fig:quadrupole_vs_dcm_vs_freq}(b)]. An additional modification of the OMCs due to the interaction between scatterers is neglected. Then, we vary $N$ from 80 to 380, estimate the time we need to solve the multiple scattering problem, compute the ring scattering cross section for each $N$, average this time after 10 iterations, and normalize it by $\tau$. Here, $\tau$ is the average time to solve a linear system with a $1000 \times 1000$ matrix and $1000 \times 1$ right-hand side using the \textit{ numpy.linalg.solve} function in Python 3.11. 

Figure~\ref{fig:time_vs_N} illustrates the results, demonstrating that the OMC reduces computational time by a factor of six, as we manage to reduce the size of the isolated-cone T-matrix from 9 for $\ell_{\rm max} = 2$ to 4 for $\ell_{\rm max} = 1$, while the error in the calculation of the ring scattering cross-section remains comparable. Moreover, using the MDA but placing the multipole center at the center of mass increases the average error from 1.42\% to 4\%. Thus, the OMC remarkably reduces the computational time within the T-matrix and multipole methods for simulating the acoustic response of discrete scatterers while maintaining a reasonable error. 

As drawbacks of the OMC approach, we have to mention first that an offset $d$ of the multipole expansion center effectively increases the radius of a sphere circumscribing the scatterer from $R_\mathrm{c}$ to $(R_\mathrm{c} + d)$. Expansion~\eqref{eq_psca} becomes then valid outside a larger circumscribing sphere,~\cite{Auguie2016May} which may lead to larger minimal spacings between the scatterers in their arrangements. Moreover, the OMC, as discussed here, has been found for a particular incidence of the external wave along the $z$-axis. As discussed in Ref.~[\onlinecite{Kildishev2023Sep}], a different illumination direction leads to different OMCs, making the OMC both frequency- and angular-dispersive [see Sec.~S5 in SM].

In conclusion, we address the problem of determining the optimal point to place the multipole expansion center in acoustic scattering. We consider individual scatterers and multi-particle structures without reflection symmetry along the $z$-axis and trace the OMC position along the axis when the scatterers are insonified by a plane wave. Our comprehensive acoustic scattering model reveals that similar to electromagnetic scattering,~\cite{Kildishev2023Sep} an optimal solution that minimizes the multipole content to produce the most efficient descriptor while retaining only a few terms of the lowest degree. We showcase the advantages of using low-degree multipoles (monopole and dipole) at this OMC, demonstrating substantial computational efficiency gains achieved in T-matrix calculations. Our approach can be extended beyond the systems described within the monopole-dipole-quadrupole approximation to systems of arbitrary geometry that require optimizing their multipole descriptions up to arbitrary multipole degrees and orders.

See the Supplementary material for the definition of scalar spherical waves (multipoles), the addition theorem and translation coefficients, the derivation of Eqs.~\eqref{eq_quadrupole_transl}-\eqref{eq_quadrupole_interpol_quartic_grad}, the analysis of the error of approximations~\eqref{eq_quadrupole_transl}, ~\eqref{eq_transl_coeffs_expansion}, and~\eqref{eq_quadrupole_interpol_quartic}, the calculation of an optimal multipole center trajectory for the rotated trimer, the calculation of an optimal multipole center for the isolated cone as a function of the angle of incidence, and the details on the computation of the acoustic T-matrices of axisymmetric objects in COMSOL Multiphysics.

\begin{acknowledgments}
The authors acknowledge Markus Nyman for helping with the numerical simulations. N.U. and C.R. acknowledge support from the Deutsche Forschungsgemeinschaft (DFG, German Research Foundation) under Germany’s Excellence Strategy via the Excellence Cluster 3D Matter Made to Order (EXC-2082/1, Grant No. 390761711) and from the Carl Zeiss Foundation via CZF-Focus@HEiKA.
\end{acknowledgments}

\section*{Data Availability Statement}
The data that support the findings of this study are openly available on GitHub at \href{https://github.com/NikUstimenko/acoustotreams.git}{https://github.com/NikUstimenko/acoustotreams.git}, Ref.~[\onlinecite{acoustotreams}].

\bibliography{main}

\end{document}



\title{Supplementary material: Optimal multipole center for subwavelength acoustic scatterers}

\author{N. Ustimenko}
\email{nikita.ustimenko@kit.edu}
\affiliation{Institute of Theoretical Solid State Physics, Karlsruhe Institute of Technology, Kaiserstrasse 12, Karlsruhe, D-76131, Germany}%

\author{C. Rockstuhl}
\affiliation{Institute of Theoretical Solid State Physics, Karlsruhe Institute of Technology, Kaiserstrasse 12, Karlsruhe, D-76131, Germany}%
\affiliation{Institute of Nanotechnology, Karlsruhe Institute of Technology, Kaiserstrasse 12, Karlsruhe, D-76131, Germany}

\author{A.V. Kildishev}
\email{kildishev@purdue.edu}
\affiliation{Elmore Family School of Electrical and Computer Engineering and Birck Nanotechnology Center, Purdue University, 1205 W State St, West Lafayette, IN 47907, USA}%


\maketitle

\section{Scalar spherical waves}
\subsection{Definition of scalar spherical waves (SSWs)}
\begin{figure}
    \centering
    \includegraphics[scale=1.75]{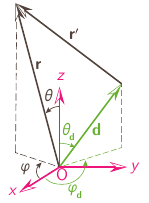}
    \caption{The radius vector of an observation point with respect to the origin ($\mathbf{r}$) and translated multipole center ($\mathbf{r}'$) connected through an offset vector $\mathbf{d} = \mathbf{r} -\mathbf{r}'$.}
    \label{fig:S1}
\end{figure}

A scalar spherical wave  (SSW, or multipole) of degree $\ell$ and order $m$ is defined as 
\begin{align}
\label{eq_S_ssw_def}
    \psi^{(n)}_{\ell, m}(k; \mathbf{r}) = z^{(n)}_{\ell}(kr) Y_{\ell m}(\theta,\varphi),
\end{align}
where $r = |\mathbf{r}|$ is the radial distance, $\theta$ is the polar angle, and $\varphi$ is the azimuthal angle as depicted in Fig.~\ref{fig:S1}. $z^{(n)}_{\ell}(kr)$ is a radial function such that $z^{(1)}_{\ell}(kr) \equiv j_{\ell}(kr)$ is a spherical Bessel function of the first kind and $z^{(3)}_{\ell}(kr) \equiv h^{(1)}_{\ell}(kr)$ is a spherical Hankel function of the first kind. In the literature, SSWs with $n = 1$ are called regular waves, which are regular at $r\to 0 \land r\to\infty$. In contrast, SSWs with $n = 3$ are referred to as singular waves, which have a singularity at $ r\to 0$, but are regular at $r\to\infty$ and obey the Sommerfeld radiation condition~\cite{schot1992}.

\subsection{Special functions}
\subsubsection{Spherical Bessel function: small-argument approximation}
\noindent To derive the dependencies of multipole coefficients on the offset, we employ the expansions of spherical Bessel functions for small argument $x \ll 1$, see \textit{e.g.} Refs.~\onlinecite{Abramowitz1972,Arfken2012}:
\begin{align}
\label{eq_S_jl_expand}
\begin{aligned}
j_0(x)&= 1 - \frac{x^2}{6} + \frac{x^4}{120} - \frac{x^6}{5040} + \mathcal{O}(x^8),\\
j_1(x)&= \frac{x}{3} - \frac{x^3}{30} + \frac{x^5}{840} + \mathcal{O}(x^7),\\
j_2(x) &= \frac{x^2}{15} - \frac{x^4}{210} + \frac{x^6}{7560} + \mathcal{O}(x^8), \\
j_3(x) &= \frac{x^3}{105} - \frac{x^5}{1890} + \mathcal{O}(x^7), \\
j_4(x) &= \frac{x^4}{945} - \frac{x^6}{20790} + \mathcal{O}(x^8)
\end{aligned}
\end{align}
\subsubsection{Spherical harmonics}
Spherical harmonics $Y_{\ell m}(\theta,\varphi)$ are defined as~\cite{Khersonskii1988}
\begin{align}
    Y_{\ell m}(\theta,\varphi) = \sqrt{\frac{2\ell + 1}{4 \pi} \frac{(\ell - m)!}{(\ell + m)!}} P^m_{\ell}(\cos \theta) \mathrm{e}^{\mathrm{i} m \varphi},
\end{align}
where $\mathrm{i}$ is the imaginary unit and the associated Legendre polynomials are given by
\begin{align}
    P^m_{\ell}(x) = \frac{(-1)^{m}}{2^{\ell}\ell!} (1-x^2)^{m/2} \frac{\mathrm{d}^{\ell + m}}{\mathrm{d}x^{\ell + m}}(x^2-1)^{\ell}.
\end{align}
where we follow Stegun's definition (see Ref.~\onlinecite{Abramowitz1972}, Ch. 8.) which includes the Condon-Shortley phase $(-1)^m$.

\subsection{Summation theorem for scalar spherical waves}
\noindent
Here, we use a summation theorem~\cite{stein1961, sack1964, Danos1965May, Kafesaki1999Nov,Wittmann2002Aug, Gonis2000, Martin2006Aug}  that expands either singular into singular or regular into regular multipoles,  
\begin{align}
\label{eq_S_regcoeff}
    \psi^{(n)}_{\red {\ell, m}}(k; \mathbf{r}) = \sum_{\ell' = 0}^{\ell_{\rm max}} \sum_{m' = -\ell'}^{\ell'} \alpha_{\blue{\ell' m'} \red{\ell m}}^{(1)}(k; \mathbf{d}) \psi^{(n)}_{\blue{\ell',m'}}(k; \mathbf{r}').
\end{align}
where $\mathbf{r}'$ and translated $\mathbf{r}$ radius vectors are connected through a translation offset vector $\mathbf{d}$ as $\mathbf{d} = \mathbf{r} -\mathbf{r}'$ [see Fig.~\ref{fig:S1}]. 

The regular translation coefficients in Eq.~\eqref{eq_S_regcoeff} read as
\begin{align}
\label{eq_translation_coeff}
\begin{aligned}
    \alpha_{\ell' m' \ell m}^{(1)}(k; \mathbf{d}) &= \sqrt{4\pi (2\ell+1)(2\ell'+1)} \ (-1)^{m} \ \mathrm{i}^{\ell'-\ell} \\
    &\cdot \sum_q \mathrm{i}^{q} \sqrt{2q+1} \psi^{(1)}_{q,m-m'}(k; \mathbf{d}) \cdot \\
    &\underbrace{\begin{pmatrix}
            \ell& \ell' & q \\
            m & -m' & m'-m
        \end{pmatrix}
        \begin{pmatrix}
            \ell& \ell' & q \\
            0& 0& 0
        \end{pmatrix}}_{\text{Wigner 3j symbols}}.
\end{aligned}
\end{align}
The finite sum is computed for $q \in \left\{\ell'+\ell, \ell'+\ell-2, ..., \max\left(|\ell'-\ell|, |m' - m| \right) \right\}$. The wave $\psi^{(1)}_{q,m-m'}(k; \mathbf{d})$ takes as argument the spherical coordinates of $\mathbf{d}$ shown in Fig.~\ref{fig:S1}.

\subsection{Derivation of Eq.~(2) from the main text}
Equation~\eqref{eq_S_regcoeff} for $\mathbf{r}' = \mathbf{r} - \mathbf{d}$ and $\ell \leq \ell_{\rm max}$ implies that the spherical waves are translated as
\begin{align}
\label{eq_S_psi_aux}
    \psi^{(3)}_{\ell',m'}(k; \mathbf{r}) = \sum_{\ell = 0}^{\ell_{\rm max}} \sum_{m = -\ell}^{\ell} \alpha_{\ell m \ell' m'}^{(1)}(k; \mathbf{d}) \psi^{(3)}_{\ell,m}(k; \mathbf{r} - \mathbf{d}).
\end{align}

The scattered pressure given by~(1) in the main text with the multipole center at $\mathbf{d} = \bm{0}$ is
\begin{align}
    p_{\rm sca}(k; \mathbf{r})  = \sum_{\ell' = 0}^{\ell_{\rm max}} \sum_{m' = -\ell'}^{\ell'} p_{\ell', m'}(k; \bm{0}) \psi^{(3)}_{\ell', m'}(k; \mathbf{r}).
\end{align}
Using~\eqref{eq_S_psi_aux} it can be rewritten as
\begin{align}
\label{eq_psca_appendix}
\begin{aligned}
    p_{\rm sca}(k; \mathbf{r})  &= \sum_{\ell = 0}^{\ell_{\rm max}} \sum_{m = -\ell}^{\ell}  \\
    &\left[ \sum_{\ell' = 0}^{\ell_{\rm max}} \sum_{m' = -\ell'}^{\ell'} \alpha_{\ell m \ell' m'}^{(1)}(k; \mathbf{d}) p_{\ell', m'}(k; \bm{0})\right] \\
    &\psi^{(3)}_{\ell, m}(k; \mathbf{r} - \mathbf{d}).
\end{aligned}
\end{align}
After comparison of Eqs.~(1) and~\eqref{eq_psca_appendix} we obtain~(2) for the translation of multipole moments:
\begin{align}
\label{eq_moments_transl}
    \boxed{p_{\red{\ell, m}}(k; \mathbf{d}) = \sum_{\ell' = 0}^{\ell_{\rm max}} \sum_{m' = -\ell'}^{\ell'} \alpha^{(1)}_{\red{\ell m} \blue{\ell' m'}}(k; \mathbf{d}) p_{\blue{\ell', m'}}(k; \bm{0})}.
\end{align}

\textit{Thus, as already written in the main text, spherical waves (multipoles) and multipole moments are translated with the same coefficients~\eqref{eq_translation_coeff} up to the permutation of the indices $\{(\ell,m),(\ell',m')\}  \rightarrow \{(\ell',m'),(\ell,m)\}$} [cf. Eqs.~\eqref{eq_S_regcoeff} and~\eqref{eq_moments_transl}].
Moreover, regular translation coefficients of SSWs~\eqref{eq_translation_coeff} with real $k$ obey the following symmetry relation~\cite{Kim2004}:
\begin{subequations}
    \begin{align}
    \alpha_{\red{\ell m} \blue{\ell' m'}}^{(1)}(k; \mathbf{d}) &= \left[\alpha_{\blue{\ell'm'} \red{\ell m} }^{(1)}(k; -\mathbf{d}) \right]^*.    
\end{align}
\end{subequations}

\section{SHIFTED ZONAL MULTIPOLES}
\subsection{Summation theorem for zonal quadrupole moment}
Truncating the summation in~\eqref{eq_moments_transl} up to $(\ell_{\mathrm{max}} = 2)$  gives an approximate three-term summation theorem for a quadrupole moment $p_{2,0}$ translated along the $z$-axis by $\mathbf{d} = d \hat{\mathbf{z}}$,
\begin{align}\label{eq_S_psi_MDQ}
    p_2(kd) \approx p_{2}(0) + \sum_{\ell' = 0}^{2}  \alpha_{\ell'}(kd) p_{\ell'}(0),
\end{align}
where $p_2(kd) \equiv p_{2,0}(k; d\hat{\mathbf{z}})$, $ p_{\ell'}(0)\equiv p_{\ell',0}(k; \bm{0})$, and $\alpha_{\ell'}(kd)\equiv\alpha^{(1)}_{2 0 \ell' 0}(k;\mathbf{d})-\delta_{\ell',2}$. Equation~\eqref{eq_S_psi_MDQ} is basically~(3) from the main text. The exact translation coefficients in~\eqref{eq_S_psi_MDQ}  are given by~\eqref{eq_translation_coeff} as
\begin{align}
\label{eq_S_transl_coeffs_rigorous}
    \begin{aligned}
        \alpha_0(kd) &= \sqrt{5} j_2(kd), \\
        \alpha_1(kd) &= -\frac{2\sqrt{3}}{\sqrt{5}}\left[j_1(kd) - \frac{3}{2}j_3(kd)\right], \\
        \alpha_2(kd) &= j_0(kd) - \frac{10}{7} j_2(kd) + \frac{18}{7} j_4(kd)-1.
    \end{aligned}
\end{align}
We further simplify~\eqref{eq_S_psi_MDQ} to convert it into a polynomial form, where we use the long-wavelength approximation (LWA) for the spherical Bessel functions from Eqs.~\eqref{eq_S_jl_expand} up to quadratic terms 
\begin{align} 
\begin{aligned}
    j_0(kd) &\approx 1 - \frac{(kd)^2}{6} + \mathcal{O}[(kd)^4], \\
    j_1(kd) &\approx \frac{kd}{3} + \mathcal{O}[(kd)^3], \\
    j_2(kd) &\approx \frac{(kd)^2}{15} + \mathcal{O}[(kd)^4], \\
    j_3(kd) &\approx 0 + \mathcal{O}[(kd)^3], \\
    j_4(kd) &\approx 0 + \mathcal{O}[(kd)^4].
\end{aligned}
\end{align}
Then we obtain the polynomial approximations of the translation coefficients [Eq.~(4) from the main text]:
\begin{align}
    \begin{aligned}\label{eq_S_alpha}
        \widetilde{\alpha}_0(kd) &= \frac{1}{3\sqrt{5}} (kd)^2, \\
        \widetilde{\alpha}_1(kd) &= -\frac{2}{\sqrt{15}} kd, \\
        \widetilde{\alpha}_2(kd) &= -\frac{11}{42}(kd)^2 .
    \end{aligned}
\end{align}
In this case,~\eqref{eq_S_psi_MDQ} reads as 
\begin{align}\label{eq_S_aprox_psi_MDQ}
    \widetilde{p}_2(kd) = p_{2}(0) + \sum_{\ell' = 0}^{2}  \widetilde{\alpha}_{\ell'}(kd) p_{\ell'}(0).
\end{align}

To test the validity of approximations~\eqref{eq_S_psi_MDQ} and~\eqref{eq_S_aprox_psi_MDQ} we introduce the relative errors of the quadrupole moment as
\begin{align}
    \mathrm{Error}_{\left[p_2\right]}\left(kd\right) &=  \frac{\left|p_2(kd)- p_2^{\mathrm{Exact}}(kd)\right|}{\left|p_2^{\mathrm{Exact}}(kd)\right|} \times 100\%,\label{eq_S_error1}\\
    \mathrm{Error}_{\left[\widetilde{p}_2\right]}\left(kd\right) &=  \frac{\left|\widetilde{p}_2(kd)- p_2^{\mathrm{Exact}}(kd)\right|}{\left|p_2^{\mathrm{Exact}}(kd)\right|}\times 100\%.\label{eq_S_error2}
\end{align}
The exact multipole moment is obtained from~\eqref{eq_moments_transl} with $\ell_{\rm max} = 5$. As multipole moments $p_{\ell'}(0)$, we consider those for a subwavelength cone with a size parameter of $\omega h/ c_0 = 0.686$.
\begin{figure}
    \centering
    \includegraphics[scale=0.58]{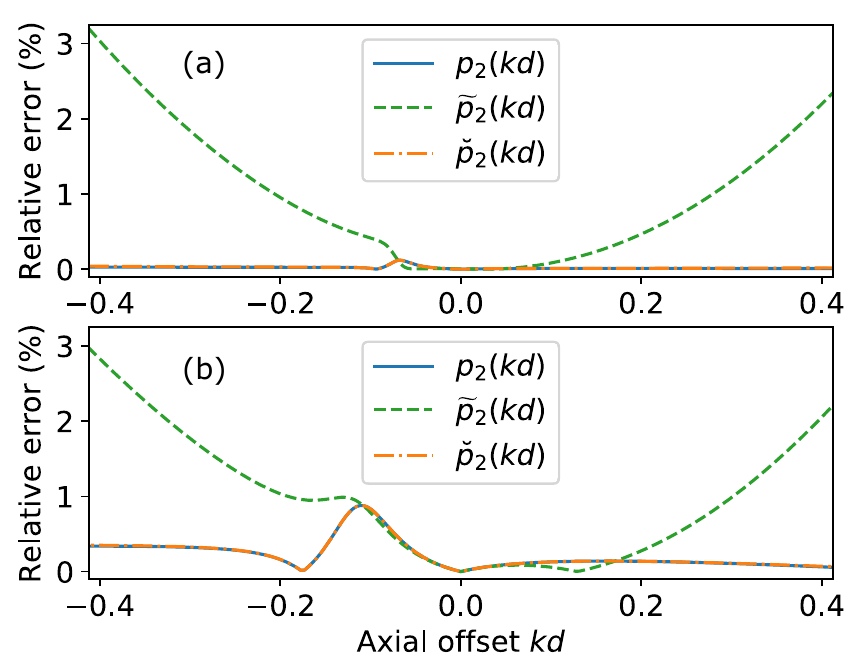}
    \caption{Relative errors~\eqref{eq_S_error1} (blue solid),~\eqref{eq_S_error2} (green dashed), and~\eqref{eq_S_error3} (orange dashed-dotted) of three-term approximations~\eqref{eq_S_psi_MDQ}, ~\eqref{eq_S_aprox_psi_MDQ}, and~\eqref{eq_S_squared_norm_2} for a cone at frequency (a) 5 kHz ($\omega h/ c_0 = 0.686$) and (b) 10 kHz ($\omega h/ c_0 = 1.371$).}
    \label{fig:error}
\end{figure}
The results are plotted in Figure~\ref{fig:error}. They manifest the applicability of the three-term approximations.

\subsection{Summation theorem for the monopole and zonal dipole moments}

Similar three-term approximations are available for the monopole and dipole multipole moments for $kd \ll 1$:
\begin{equation}
\begin{aligned}
        p_{0}(kd) &\approx j_0(kd) p_{0} + \sqrt{3} j_1(kd) p_{1} + \sqrt{5} j_2(kd) p_{2}, \\
        p_{1}(kd) &\approx -\sqrt{3} j_1(kd) p_{0} +  \left[ j_0(kd) - j_2(kd)\right] p_{1} \\
        &\quad - \alpha_1(kd) p_{2},
\end{aligned}
\end{equation}
where $\alpha_1(kd)$ is defined by~\eqref{eq_S_transl_coeffs_rigorous}, which give the LWA series expansions: 
\begin{equation}
\begin{aligned}
\label{eq_S_a00_a10_interpol}
        \widetilde{p}_{0}(kd) &\approx p_{0} + \frac{\sqrt{5}}{2} b_{1,k} kd  + \frac{1}{3\sqrt{5}} b_{0,k} (kd)^2 , \\
        \widetilde{p}_{1}(kd)&\approx p_{1} - \sqrt{15}\left(\frac{83}{210} b_{0,k} + b_{2,k} \right) kd  \\
        &\quad + \frac{7}{4 \sqrt{15}} b_{1,k} (kd)^2,
\end{aligned}   
\end{equation}
where coefficients $b_{i,k}$ are defined in~\eqref{eq_S_bi_def}.

\subsection{The squared norm of shifted zonal quadrupole moment}
\subsubsection{Quadratic approximation: real \textit{k}}
We rewrite Eq.~\eqref{eq_S_aprox_psi_MDQ} as a Taylor expansion in $kd \equiv kd$: 
\begin{align}
\label{eq_p2_tilde}
\begin{aligned}
    \widetilde{p}_{2}(kd) &= b_{0,k} + b_{1,k} (kd) + b_{2,k} (kd)^2,
\end{aligned}
\end{align}
where the coefficients are functions of frequency:
\begin{equation}
\begin{aligned}\label{eq_S_bi_def}
        b_{0,k} &= p_{2},  b_{1,k} = -\frac{2}{\sqrt{15}}p_{1},\\ &\text{and }\\
        b_{2,k} &= \frac{1}{3\sqrt{5}}p_{0} - \frac{11}{42}p_{2}.   
\end{aligned}
\end{equation}

The squared norm of the shifted zonal quadrupole moment is given by~(5) from the main text:
\begin{align}
\begin{aligned}\label{eq_S_squared_norm_1}
|\widetilde{p}_{2}(kd)|^2 = f_{0,k} + f_{1,k} (kd) + f_{2,k} (kd)^2,
\end{aligned}
\end{align}
where
\begin{equation}
\begin{aligned}
    \label{eq_S_fi_def}
        f_{0,k}&= |b_{0,k}|^2,\, f_{1,k}  = 2\Re\left[b_{0,k}^* b_{1,k}\right]\\ 
        f_{2,k}&= |b_{1,k}|^2 + 2\Re\left[b_{0,k}^* b_{2,k}\right].
\end{aligned}
    \end{equation}

\subsubsection{Quadratic approximation: complex \textit{k}}
Let us consider the case of complex $k = \Re[k] + \ii\Im[k]$ in~\eqref{eq_p2_tilde} and calculate the squared norm as $\widetilde{p}_{2}^* \cdot \widetilde{p}_{2}$, then:
\begin{align}
    |\widetilde{p}_{2}(k; d)|^2 = f_{0,k} + f_{1,k} d + f_{2,k} d^2 ,
\end{align}
where $f_{0,k}$ is the same as in~\eqref{eq_S_fi_def}, while $f_{1,k}$ and $f_{2,k}$ read as:
\begin{equation}
\begin{aligned}
\label{eq_S_fi_def_complexk}
 f_{1,k}  &= 2 \left( \Re[k]\Re\left[b_{0,k}^* b_{1,k}\right] -   \Im[k]\Im\left[b_{0,k}^* b_{1,k}\right] \right), \\
 f_{2,k}  &= |k|^2 |b_{1,k}|^2 + 2 \left(
\Re\left[k^2 \right] \Re\left[b_{0,k}^* b_{2,k}\right] \right.\\
&\left.\qquad\qquad- \Im\left[k^2 \right]\Im\left[b_{0,k}^* b_{2,k}\right]\right). 
\end{aligned}   
\end{equation}

\subsubsection{Quartic approximation: real \textit{k}}
To present the quadrupole moment as a polynomial with powers of $(kd)$ up to fourth, we use the corresponding expansions of spherical Bessel functions~\eqref{eq_S_jl_expand}, giving more accurate approximations of the translation coefficients:  
\begin{align}
\label{eq_S_transl_coeffs_expansion_quartic}
    \begin{aligned}
        \breve{\alpha}_0(kd) &= \frac{1}{3\sqrt{5}} \left[ (kd)^2 - \frac{(kd)^4}{14}  \right], \\
        \breve{\alpha}_1(kd) &= -\frac{2}{\sqrt{15}} \left[ kd - \frac{(kd)^3}{7}\right], \\
        \breve{\alpha}_2(kd) &= -\frac{11}{42}(kd)^2 + \frac{1}{56}(kd)^4.
    \end{aligned}
\end{align}
By inserting~\eqref{eq_S_transl_coeffs_expansion_quartic} into~\eqref{eq_S_psi_MDQ}, we obtain the desired expansion of the quadrupole moment:
\begin{align}
\begin{aligned}
    \breve{p}_{2}(kd) &= \sum\limits_{n=0}^4 b_{n,k}(kd)^n ,
\end{aligned}
\end{align}
where
\begin{align}
\label{eq_S_bi_def_2}
        b_{3,k} = \frac{2}{7 \sqrt{15}}p_{1}, \, b_{4,k} = \frac{1}{56}p_{2} -\frac{1}{42\sqrt{5}}p_{0},
\end{align}
while $b_{0,k}$, $b_{1,k}$, and $b_{2,k}$ are the same as in~\eqref{eq_S_bi_def}. 

For the quartic approximation, we also introduce the relative errors of the quadrupole moment, 
\begin{align}
    \mathrm{Error}_{\left[\breve{p}_2\right]}\left(kd\right) &=  \frac{\left|\breve{p}_2(kd)- p_2^{\mathrm{Exact}}(kd)\right|}{\left|p_2^{\mathrm{Exact}}(kd)\right|} \times 100\% \label{eq_S_error3}
\end{align}
Figure~\ref{fig:error}  compares $\mathrm{Error}_{\left[\breve{p}_2\right]}$  with  $\mathrm{Error}_{\left[p_2\right]}\left(kd\right)$ and $\mathrm{Error}_{\left[\widetilde{p}_2\right]}$ , obtained above from Eqs. \eqref{eq_S_error1} and \eqref{eq_S_error2}, respectively. The relative error achievable for $\breve{p}_2$ is much lower than the error of $\widetilde{p}_2$ and is almost indiscernible from the error of $p_2$, calculated with a truncated three-term addition theorem with exact translation coefficients.

Thus, for the squared norm, we have:
\begin{align}
\begin{aligned}\label{eq_S_squared_norm_2}
|\breve{p}_{2}(kd)|^2 &= \sum\limits_{n=0}^4 f_{n,k}(kd)^n,
\end{aligned}
\end{align}
where
\begin{align}
\label{eq_S_fi_def_2}
    \begin{aligned}
        f_{3,k} &= 2\Re\left[b_{0,k}^* b_{3,k} + b_{1,k}^* b_{2,k}\right], \\ 
        f_{4,k} &= \left|b_{2,k}\right|^2 +2\Re\left[b_{0,k}^* b_{4,k} + b_{1,k}^* b_{3,k}\right]. 
    \end{aligned}
\end{align}
while $f_{0,k}$, $f_{1,k}$, and $f_{2,k}$ are the same as in~\eqref{eq_S_fi_def}. In the case of a complex $k$, the coefficients $f_{3,k}$ and $f_{4,k}$ can be derived as in~\eqref{eq_S_fi_def_complexk}.
\begin{figure}[h!]
    \centering
    \includegraphics[scale=0.55]{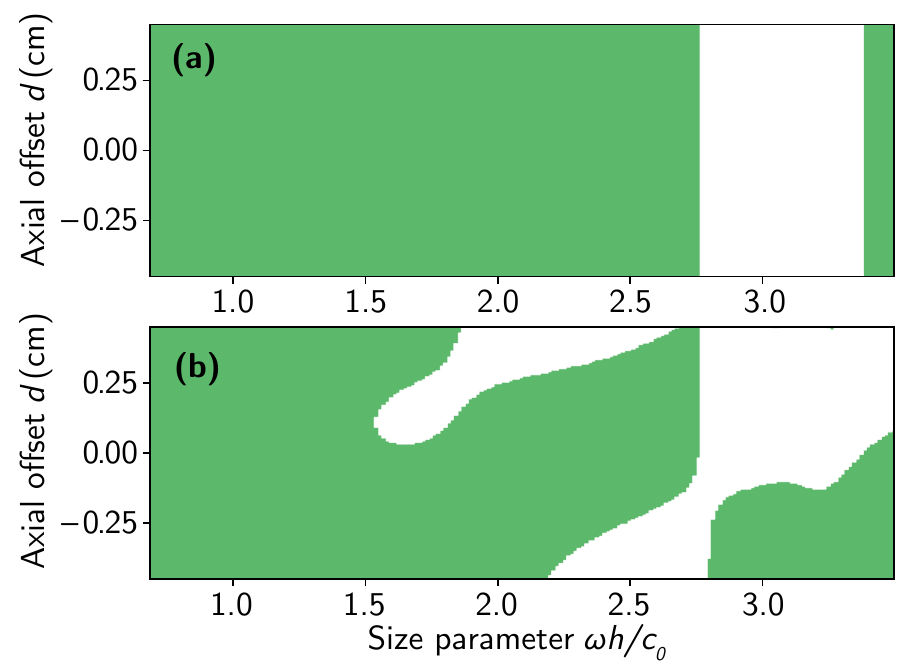}
    \caption{Second derivative of (a) quadratic~\eqref{eq_S_squared_norm_1} and (b) quartic~\eqref{eq_S_squared_norm_2} interpolations of the quadrupole moment of the cone. The green color corresponds to positive values, while the white to negative ones.}
    \label{fig:sd}
\end{figure}

\subsubsection{Sextic approximation: real \textit{k}}
Here, we present the sextic approximation of the quadrupole moment to calculate the OMC at the monopole resonance of the cone. In the sextic approximation, the translation coefficients~\eqref{eq_S_transl_coeffs_rigorous} read as
\begin{align}
\label{eq_S_transl_coeffs_expansion_sextic}
    \begin{aligned}
        \widehat{\alpha}_0(kd) &= \frac{1}{3\sqrt{5}} \left[ (kd)^2 - \frac{(kd)^4}{14} + \frac{(kd)^6}{504}  \right], \\
        \widehat{\alpha}_1(kd) &= -\frac{2}{\sqrt{15}} \left[ kd - \frac{(kd)^3}{7} + \frac{(kd)^5}{168}\right], \\
        \widehat{\alpha}_2(kd) &= -\frac{11}{42}(kd)^2 + \frac{1}{56}(kd)^4 - \frac{17}{33264}(kd)^6.
    \end{aligned}
\end{align}
They provide us with the following approximation of the quadrupole moment
\begin{align}
\begin{aligned}
    \widehat{p}_{2}(kd) &= \sum\limits_{n=0}^6 b_{n,k}(kd)^n,
\end{aligned}
\end{align}
where
\begin{align}
    b_{5,k} = -\frac{1}{84\sqrt{15}}p_1, b_{6,k} = \frac{1}{1512\sqrt{5}}p_0 - \frac{17}{33264}p_2,
\end{align}
while $b_{0,k}$, $b_{1,k}$, and $b_{2,k}$ are the same as in~\eqref{eq_S_bi_def}; $b_{3,k}$, and $b_{4,k}$ are the same as in~\eqref{eq_S_bi_def_2}.

The squared norm is then
\begin{align}
\label{eq_S_squared_norm_3}
    |\widehat{p}_{2}(kd)|^2 &= \sum\limits_{n=0}^6 f_{n,k}kd^n,
\end{align}
where
\begin{align}
\begin{aligned}
    f_{5,k} &= 2\Re\left[b_{0,k}^* b_{5,k} + b_{1,k}^* b_{4,k}+ b_{2,k}^* b_{3,k}\right], \\ 
    f_{6,k} &= \left|b_{3,k}\right|^2 +2\Re\left[b_{0,k}^* b_{6,k} + b_{1,k}^* b_{5,k} + b_{2,k}^* b_{4,k}\right],
\end{aligned}
\end{align}
and the other coefficients are the same as in~\eqref{eq_S_squared_norm_2}.


\begin{figure}
    \centering
    \includegraphics[scale=0.5]{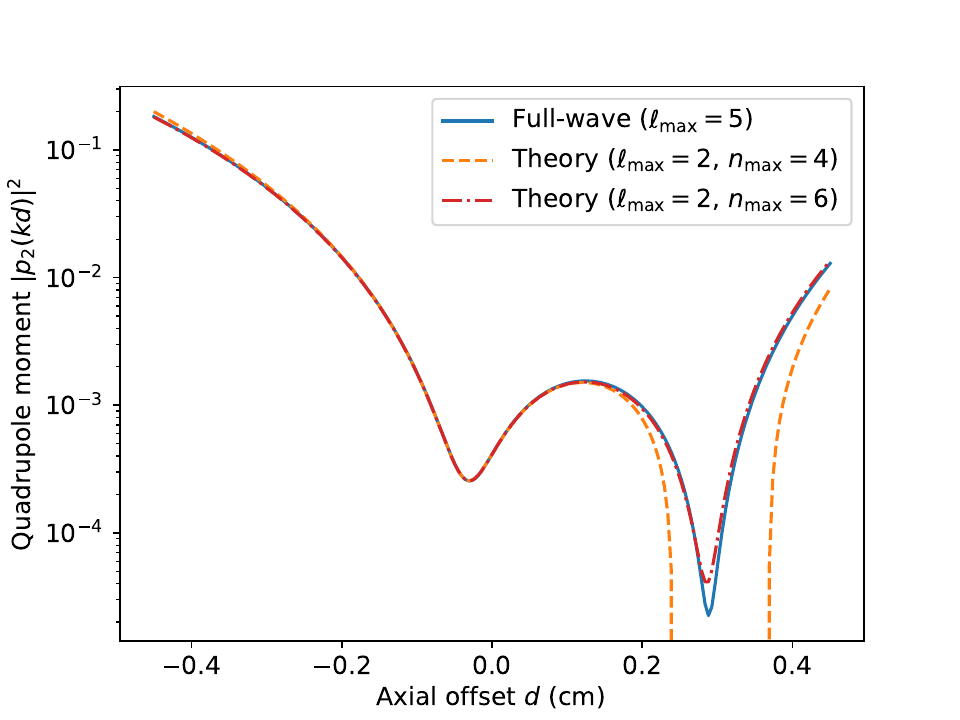}
    \caption{The squared amplitude of the cone zonal quadrupole moment as a function of $d$ at the monopole resonance on a logarithmic scale (solid blue). The dashed orange and dashed-dotted red curves show its quartic $|\breve{p}_2|^2$ ($n_{\rm max} = 4$) and sextic $|\widehat{p}_2|^2$ ($n_{\rm max} = 6$) approximations given by~\eqref{eq_S_squared_norm_2} and~\eqref{eq_S_squared_norm_3}, respectively.}
    \label{fig:omc_mr}
\end{figure}

\section{Optimal multipole center}
\subsection{Linear equation}
\noindent An optimal offset $d_{\rm opt}$ is a solution to $\partial_{kd}|\widetilde{p}_{2}(kd)|^2 = 0$, \textit{i.e.} a real root of the linear equation that stems from~\eqref{eq_S_squared_norm_1},
\begin{align}
\label{eq_S_dopt_linear}
     f_{1,k} + 2 f_{2,k} (kd) = 0.
\end{align}
The solution is a minimum when the second derivative $\partial^2_{kd}|\widetilde{p}_{2}(kd)|^2$ is positive, which implies $f_{2,k} > 0$. Figure~\ref{fig:sd}(a) plots $f_{2,k}$ as a function of the size parameter $\omega h/c_0$ for the cone. Clearly, the quadratic approximation~\eqref{eq_S_squared_norm_1} is valid for low frequencies $\omega h/c_0 < 2.75$.
\subsection{Cubic equation}
\noindent
An optimal offset $d_{\rm opt}$ is a solution to $\partial_{kd}|\breve{p}_{2}(kd)|^2 = 0$, \textit{i.e.}, a real root of the cubic equation, obtained from~\eqref{eq_S_squared_norm_2} as,
\begin{align}
\label{eq_S_dopt_cubic}
     f_{1,k} + 2 f_{2,k} (kd) + 3 f_{3,k} (kd)^2 + 4 f_{4,k} (kd)^3 = 0.
\end{align}
The second derivative $\partial^2_{kd}|\breve{p}_{2}(kd)|^2$ given by $(2 f_{2,k} + 6 f_{3,k} d_{k} + 12 f_{4,k} d_{k}^{2})$ is plotted in Fig.~\ref{fig:sd}(b).
Approximation~\eqref{eq_S_squared_norm_2} is applicable for all frequencies considered except a narrow band around $\omega h / c_0 = 2.75$ when the minimum turns into a maximum [see Figure~2 in the main text].\footnote{Although other polynomial forms for $\partial_{kd}|p_2(kd)|^2$, such as quadratic or quartic, are available for the respective cubic and quintic approximations of $|p_2(kd)|^2$, we do not show them here for brevity.
}

\begin{figure}
    \centering
    \includegraphics[scale=0.47]{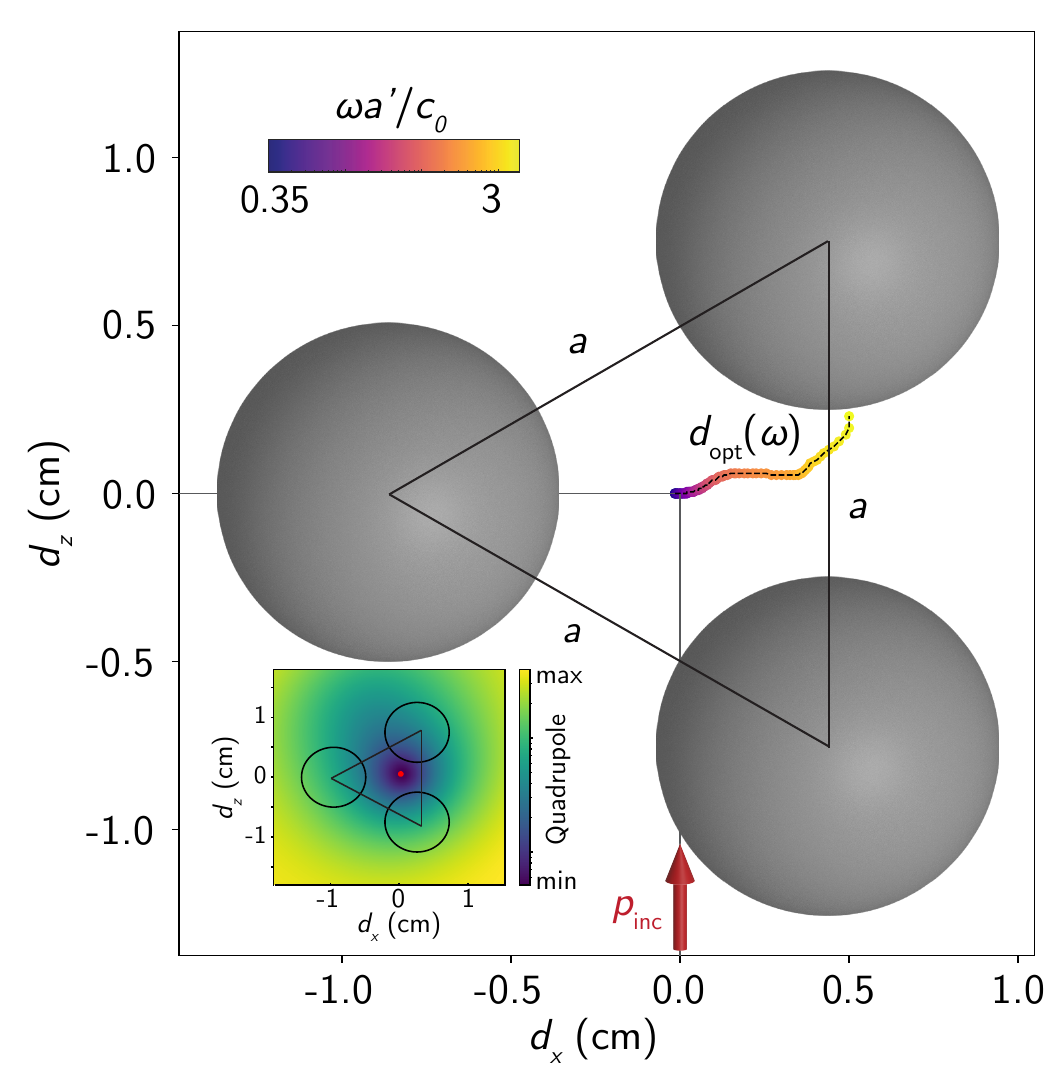}
    \caption{Positions of the optimal multipole center shown by markers as a function of frequency from 1 kHz (blue, $\omega a'/c_0 = 0.343$) to 11 kHz (yellow, $\omega a'/c_0 = 3.77$). The effective size $a'=a+2R=5R$. The center of mass is located at $\mathbf{d} = \bm{0}$. The inset shows the global quadrupole moment as a function of $\mathbf{d}$ at 8 kHz ($\omega a'/c_0 = 1.1$), whose minimum is marked with a red dot.}
    \label{fig:dopt}
\end{figure}

\subsection{Quintic equation}
An optimal offset $d_{\rm opt}$ is a solution to $\partial_{kd}|\widehat{p}_{2}(kd)|^2 = 0$, \textit{i.e.}, a real root of the quintic equation, obtained from~\eqref{eq_S_squared_norm_3} as,
\begin{align}
\label{eq_S_dopt_quintic}
\sum_{n=1}^6 nf_{n,k} (kd)^{n - 1} = 0.
\end{align}
This equation can be used, for example, to predict the OMC position at the monopole resonance of the cone. The quartic model is not applicable in this case because $|\breve{p}_2|^2$ takes negative values in the vicinity of the global minimum $d \approx 0.29$ cm [see Figure~\ref{fig:omc_mr}].

\begin{widetext}
\begin{figure*}
    \centering
    \includegraphics[scale=0.47]{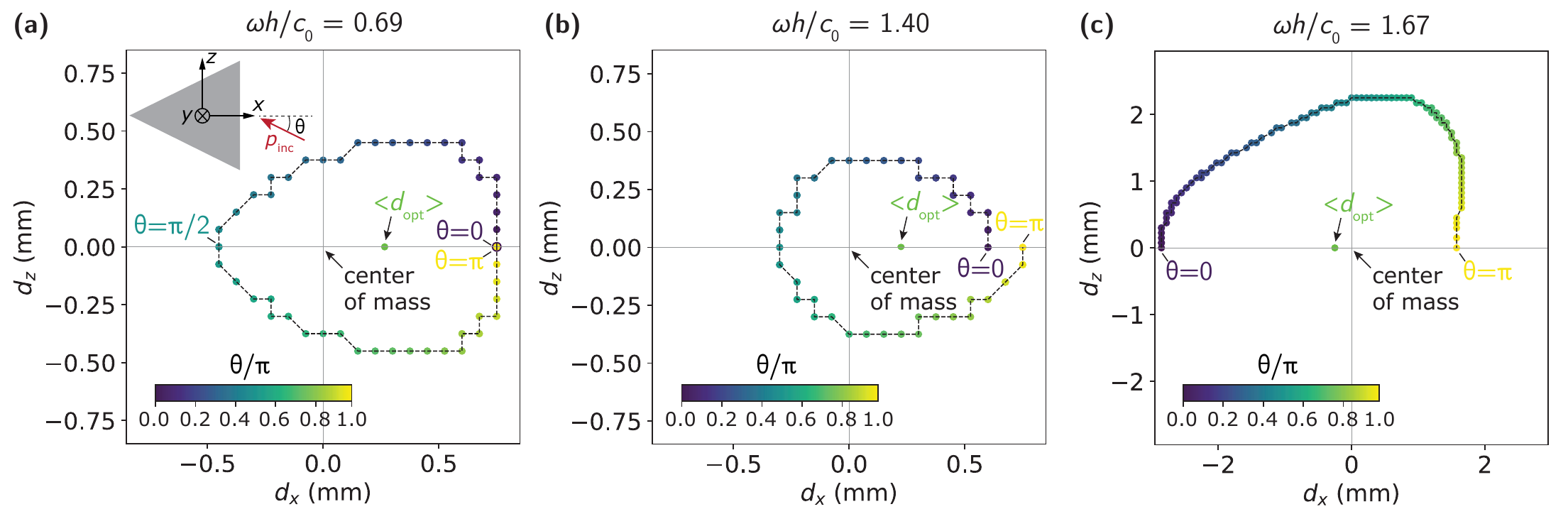}
    \caption{Optimal multipole center position in the $xz$ plane for the cone as a function of the angle of incidence $\theta$ encoded by color. The inset illustrates a definition of $\theta$. The frequency is 18.9 kHz ($\omega h/c_0 = 2.59$). The green marker shows $d_{\rm opt}$ averaged over $\theta$ and $\varphi$.}
    \label{fig:theta}
    \includegraphics[scale=0.395]{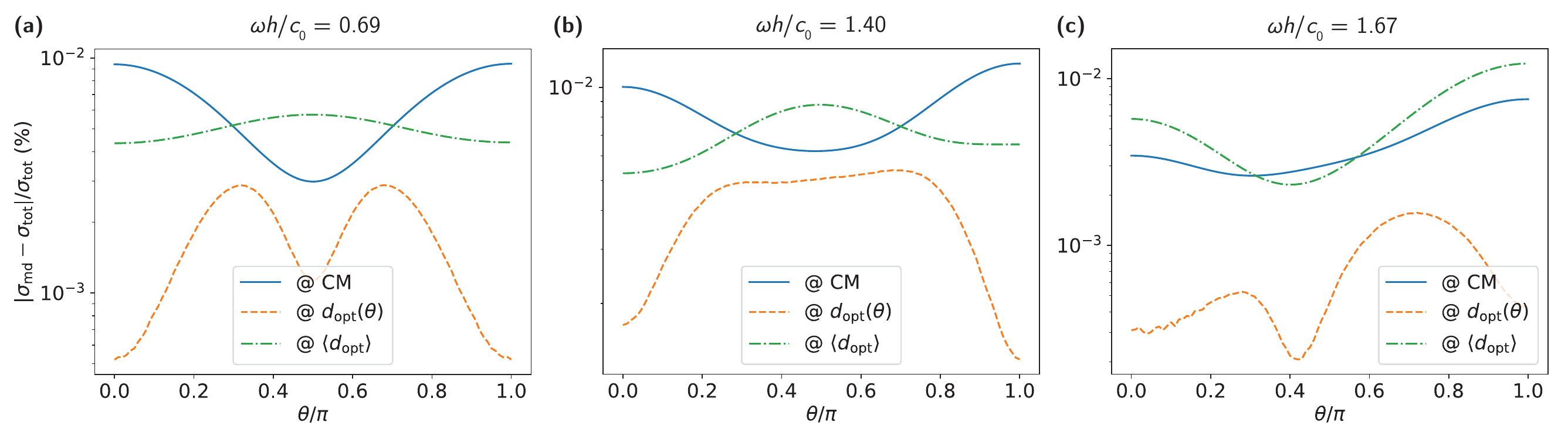}
    \caption{Relative error of the cone scattering cross section calculated in the MDA as a function of $\theta$. The frequency is 18.9 kHz ($\omega h/c_0 = 2.59$).}
    \label{fig:theta_error}
\end{figure*}
\end{widetext}

\section{Optimal multipole center for the rotated trimer}
In the main text, we consider an equilateral trimer of identical spheres of $C_{2v}$ symmetry axis coinciding with the direction of plane-wave insonification, the $z$-axis. Here, we turn to the same trimer rotated in the $xz$-plane around the $y$-axis by $\pi/2$, so that its $C_{2v}$ symmetry axis coincides with the $x$-axis, while the incident wave still impinges along the $z$-axis [see Fig.~\ref{fig:dopt}].
In the latter case, the planar symmetry with respect to the $xz$-plane is the only remaining constraint, and we need to consider vector offsets $\mathbf{d} = (d_x, 0, d_z)$ to find an optimal multipole center of such a rotated trimer. The results are presented in Figure~\ref{fig:dopt}. For small frequencies, the OMC coincides with the center of mass, as it does in the axisymmetric case, moving for some time solely along the $x$-axis until the retardation effects remain insignificant; this horizontal shift of the OMC is due to the dominant unbalanced interaction between the single sphere to the left of the CM with two vertically aligned spheres to the right. For higher frequencies, the OMC moves up towards the upper sphere and stabilizes at $d_x \approx 0.5$ cm. This effect reflects retardation that results in an increasing phase mismatch between the vertically aligned spheres. \footnote{We have to comment here that at the higher frequencies, the LWA loses its accuracy, and the higher-up multipole moments (beyond MDQA) should be considered.} 

\section{Angular dependence of the cone optimal multipole center}
Until this section, we have been considering the normal incidence of the external plane wave. Here, we study the angular dependence of the optimal multipole center at a fixed frequency.
Figure~\ref{fig:theta} shows the OMC for the rotated cone in the $xz$ plane as a function of the angle $\theta \in [0, \pi]$ that determines the wavevector of the incident wave as $\mathbf{k} = -k \cos \theta \cos \varphi \hat{\mathbf{x}} -k \cos \theta \sin \varphi \hat{\mathbf{y}} + k \sin \theta \hat{\mathbf{z}}$ as shown by the inset in Fig.~\ref{fig:theta}(a). We assume $\varphi = 0$ because for $\varphi \neq 0$ the OMC can be obtained by rotating the vector around the $z$ axis: $d_x \hat{\mathbf{x}} \to d_x\cos \varphi \hat{\mathbf{x}} + d_x\sin \varphi \hat{\mathbf{y}}$. In Figure~\ref{fig:theta} the OMC is on the $z$ axis for $\theta = 0$ or $\theta = \pi$  that matches the previous result in Figure~2 in the main text. The variation of $\theta$ leads to the motion of the OMC position that is also frequency-dependent. For example, at $\omega h / c_0 \lesssim 1$, the OMCs for $\theta = 0$ and $\theta = \pi$ coincide, and the OMC is also symmetric with respect to $\theta \to (\pi/2 - \theta)$  [see Fig.~\ref{fig:theta}(a)]. For higher frequencies, the symmetry is getting broken [see Fig.~\ref{fig:theta}(b)], while at the monopole resonance, the OMC is very sensitive to the direction of incidence. 
Figure~\ref{fig:theta} also shows values of the OMC coordinates averaged over $\theta$ and $\varphi$. This value can be used as an angle-independent OMC at the considered frequency, providing a lower error of the scattering cross-section in the MDA than the center of mass in the paraxial regime in the nonresonant cases [see Fig.~\ref{fig:theta_error}]. At the same time, the angle-dependent OMC provides lower error across the entire range of angles of incidence.

\section{Calculation of the acoustic T-matrix in COMSOL Multiphysics}
\begin{figure}[h!]
    \centering
    \includegraphics[scale=0.35]{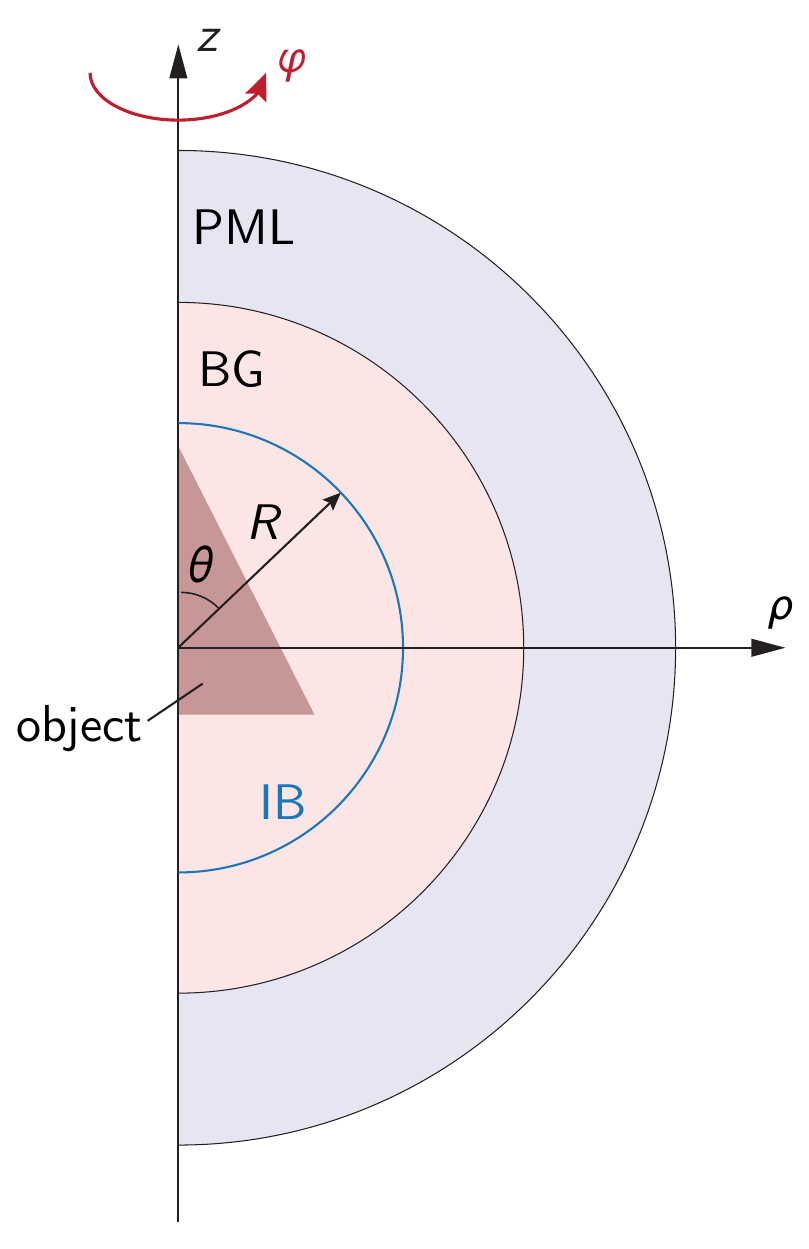}
    \caption{2D axisymmetric model in COMSOL to compute the acoustic T-matrix of an object. The model also includes a background medium (BG), an integration boundary (IB), and a perfectly matched layer (PML).}
    \label{fig:geom_comsol}
\end{figure}

For objects with cylindrical rotational symmetry, we can calculate the T-matrix in 2D axisymmetric geometry in COMSOL Multiphysics. Figure~\ref{fig:geom_comsol} presents a sketch of the system, which consists of the object, the background medium, and the perfectly matched layer (PML). The PML is used to truncate the computational domain, being an artificial absorbing domain to avoid any reflected waves. All domains belong to the Pressure Acoustics module if the object supports no shear waves; otherwise, the object domain belongs to the Solid Mechanics module. 

By definition, a column of the T-matrix contains the scattered pressure field coefficients $p_{\ell_{\rm out},m_{\rm out}}$,
\begin{align}
    p_{\rm sca}(k; \mathbf{r}) = \sum_{\ell_{\rm out} = 0}^{\ell_{\rm max}} \sum_{m_{\rm out} = -\ell_{\rm out}}^{\ell_{\rm out}} p_{\ell_{\rm out},m_{\rm out}}(k) \psi^{(4)}_{\ell_{\rm out},m_{\rm out}}(k; \mathbf{r})\,,
\end{align}
while the background field is a regular multipole field $p_{\rm inc}(k; \mathbf{r}) = \psi^{(1)}_{\ell_{\rm in},m_{\rm in}}(k; \mathbf{r})$. Since the time dependence in COMSOL is given by $\mathrm{e}^{\mathrm{i}\omega t}$, we use different singular multipoles $\psi^{(4)}_{\ell_{\rm out},m_{\rm out}}(k; \mathbf{r}) = h^{(2)}_{\ell_{\rm out}}(kr) Y_{\ell_{\rm out},m_{\rm out}}(\theta, \varphi)$ with a spherical Hankel function of the second kind. The desired coefficients, which meet $\mathrm{e}^{-\mathrm{i}\omega t}$ time evolution, can be then calculated as the integrals over the boundary of radius $R$,
\begin{align}
   p^*_{\ell_{\rm out},m_{\rm out}}(k) = \frac{2 \pi}{R h^{(2)}_{\ell_{\rm out}}(kR)}\int\limits_{0}^{\pi} \mathrm{d}l \ \sin\theta \ p_{\rm sca}(k; \mathbf{r}) Y_{\ell_{\rm out},m_{\rm out}}(\theta, 0),
\end{align}
where $^*$ stands for the complex conjugation, $\mathrm{d}l = R \mathrm{d} \theta$ and $\mathbf{r} = R(\sin \theta, 0, \cos \theta)$. The key feature of axisymmetric geometry is the conservation of the projection of angular momentum $m_{\rm in} = m_{\rm out}$. Hence, one can only make parametric sweeps of $\ell_{\rm out}$ and $\ell_{\rm in}$ in the range of $0$ to $\ell_{\rm max}$, and an auxiliary sweep of $m_{\rm in}$ the range of $-\min(\ell_{\rm out}, \ell_{\rm in})$ to $\min(\ell_{\rm out}, \ell_{\rm in})$.

The computed T-matrix as a function of frequency is then stored in the .hdf5 format and exported to our open-source code \textit{acoustotreams} to perform further simulations. A few examples of acoustic scattering calculations using this code are available as well~\cite{acoustotreams}. Although the T-matrix provides a sufficient description of the scattering, the scattering cross-section and multipole decomposition can be computed directly in COMSOL using methods described in Refs.~\onlinecite{Tsimokha2022Apr} and~\onlinecite{Deriy2022Feb}.

\APLsep
\bibliography{main}